# Xylem embolism and bubble formation during freezing suggest complex dynamics of pressure in *Betula pendula* stems

Katline Charra-Vaskou[1], Anna Lintunen[2,3], Thierry Améglio[1], Eric Badel[1], Hervé Cochard[1], Stefan Mayr[4], Yann Salmon[2,3], Heikki Suhonen[5], Mahaut van Rooij[1], Guillaume Charrier[1]*

[1] Université Clermont Auvergne, INRAE, PIAF, 63000 Clermont-Ferrand, France

[2] Institute for Atmospheric and Earth System Research / Physics, Faculty of Science, University of Helsinki, Finland

[3] Institute for Atmospheric and Earth System Research / Forest Science, Faculty of Agriculture and Forestry, University of Helsinki, Finland

[4] Institute for Botany, University of Innsbruck, Austria

[5] Department of Physics, University of Helsinki, Finland

*: corresponding author

Email: guillaume.charrier@inrae.fr

Tel: +33 4 43 76 14 21

UMR PIAF, INRAE Site de Crouel

5, chemin de Beaulieu

63000 Clermont-Ferrand

Running title

Bubble formation during freezing in *Betula pendula*

Highlight

High resolution computed tomography observations of freezing stems suggested that bubble formation is driven by the complex dynamic of pressure during freezing, affecting their likelihood to expand at thawing.

Keywords

Acoustic emissions, Freezing stress, Gas bubbles, High resolution computed tomography, Plant hydraulics, Pressure, Tension, Xylem anatomy

# Abstract

Freeze-thaw-induced embolism is a key limiting factor for perennial plants in frost-exposed environments. Gas bubbles are formed during freezing and expand during thawing resulting in xylem embolism. However, when water freezes, its volume increases by 9%, generating local pressures, which can affect the formation of bubbles. To characterize local dynamic of pressure-tension and physical state of the sap during freeze-thaw cycles, we simultaneously used ultrasonic acoustic emissions analysis and synchrotron-based high resolution computed tomography on the diffuse-porous species *Betula pendula*. Visualization of individual air-filled vessels was performed to measure freeze-thaw induced embolism after successive freeze-thaw cycles down to -10°C or -20°C during the leafy and the leafless periods. We also measured the distribution of gas bubbles in frozen xylem of *Betula pendula*, and made additional continuous monitoring of embolism spreading during one freeze-thaw cycle using a dedicated cooling system that allowed X-ray scanning during freezing and thawing. Experiments confirmed that ultrasonic emissions occurred after the onset of ice formation, together with bubble formation, whereas the development of embolism took place after thawing in all cases. The pictures of frozen tissues indicated that the positive pressure induced by the volumetric increase of ice can provoke inward flow from the cell wall toward the lumen of the vessels. We found no evidence that wider vessels within a tissue were more prone to embolism although the occurrence of gas bubbles in larger conduits would make them prone to earlier embolism. These results highlight the need to monitor local pressure as well as ice and air distribution during xylem freezing to understand the mechanism leading to frost-induced embolism.

# Introduction

Maintaining the hydraulic soil-plant-atmosphere continuum is essential for plants. In vascular plants, the xylem is the main pathway for water transport from the roots to the leaves. However, embolism in xylem conduits can results in partial or total loss of hydraulic conductivity due to tension in the xylem sap. Among abiotic environmental factors inducing hydraulic failure, frost exposure is particularly relevant at the cold edge of plant distribution (Koerner, 1998; Charrier *et al.*, 2013) and can lead to high levels of xylem embolism in trees (Mayr *et al.*, 2007; Charrier *et al.*, 2014a; Mayr and Améglio, 2016). According to the classical "thaw-expansion hypothesis" (Mayr & Sperry, 2010), the mechanism lies in two main steps: first, when xylem sap turns into ice, gas bubbles, which are less soluble in solid than in liquid sap, are entrapped in the ice lattice in the center of the *lumina* of hydraulic conduits (Sucoff, 1969; Ewers, 1985; Robson *et al.*, 1988). Second, on thawing, the bubbles may expand and embolize the whole conduit depending on the bubble size and the pressure of the surrounding sap (Lemoine *et al.*, 1999; Hacke and Sperry, 2001, Charrier *et al.*, 2014a, Charra-Vaskou *et al.*, 2016). According to this theory, bigger bubbles would form in wider vessels and would expand at lower tension than smaller ones, according to Laplace's law; making wide vessels more vulnerable to embolism (Améglio *et al*., 2002; Cruiziat *et al.*, 2002; Pittermann and Sperry, 2003; 2006). The distribution of gas bubbles formed during ice formation is in general (in free water) also affected by the rate of freezing, the concentration of dissolved air, the pressure of the system, the thickness of the layer of water ahead of the growing ice and the escape of bubbles by buoyancy (Carte *et al.*, 1962). These factors likely play a role also in the formation of gas bubbles in freezing xylem (Sperry & Sullivan, 1992).

Inter-specific studies reported that species with wider xylem conduits are more prone to embolism (Davis *et al.*, 1999; Feild and Brodribb 2001; Pitterman and Sperry, 2003; 2006; Stuart et al. 2007; Choat et al. 2011; Charrier *et al.*, 2013; 2014a). Accordingly, in ecology and evolutionary biology, vessel diameter has successfully been used as a proxy for freezing tolerance to explain plant species distribution (Olson *et al.*, 2018, Zanne *et al.*, 2018). However, the occurrence of acoustic emissions during the freezing process (implying release of tension from the cell walls), a significant effect of minimum temperature during the freeze-thaw cycle on embolism formation and the increase of embolism formation after several successive freeze-thaw cycles suggest that the prevailing thaw expansion hypothesis might not be enough to explain the gas dynamics during the freezing process of wood.

Ultrasonic acoustic emissions (UEs) have been recorded during embolism formation in the xylem under summer drought (Pena and Grace, 1986; Salleo *et al.*, 2000; Jackson and Grace, 1996; Rosner *et al.*, 2006; Mayr and Rosner, 2011) and frost stress (Raschi *et al.*, 1989; Kikuta and Richter, 2003; Mayr and Zublasing, 2010). Acoustic waves are generated in the wood during the freezing process, suggesting that a sudden release of energy triggers a vibration of the cell walls (Mayr *et al.*, 2003; Mayr *et al.*, 2007). Two independent stages in acoustic activity in the wood during freezing have been identified: (i) at the onset of freezing and during ice propagation and (ii) after latent heat dissipation (Charrier *et al.*, 2015). While the first stage is tightly related to the ice nucleation (breakdown of the metastability of the supercooled sap) and ice propagation (Charrier *et al.*, 2015), the second stage could be related to the displacement of air-water menisci in the pits (Charrier *et al.*, 2014a). Recent studies therefore suggested that frost-induced UEs would be generated by the energy relaxation of cavitation events in the freezing sap ("freeze-cavitation hypothesis"; Mayr *et al.*, 2007; Kasuga *et al.*, 2015; Charra-Vaskou *et al.*, 2016; Charrier *et al.*, 2017).

During freezing, the very low chemical potential of ice (-1.16 MPa per Kelvin below the freezing point; Hansen and Beck, 1988) attracts water molecules towards the solid-liquid interface. The sap solutes are expelled from the ice lattice towards the remaining liquid water (Sevanto *et al.*, 2012), that creates a tension high enough to reach the cavitation threshold in the pits, thus inducing pit aspiration (Maruta *et al.*, 2022) and air seeding from one conduit to the adjacent conduit (Charrier *et al.*, 2014a). Low minimum temperature during a freezing process thus induces higher embolism on thawing (Ball *et al.*, 2006; Charrier *et al.*, 2014a), because the lower the temperature, the lower the chemical potential of ice. This mechanism could also explain why successive freeze-thaw cycles result in higher embolism formation, as more conduits after each freeze-thaw cycle can act as an air source for embolism to spread and water could be locally redistributed, creating water potential gradients (Ball *et al.*, 2006; Charra-Vaskou *et al.*, 2016; Charrier *et al.*, 2017).

Direct measurements of gas bubble characteristics in frozen xylem are rare (although see Sucoff, 1969; Robson *et al.*, 1988). A recent study with high resolution computed (X-ray) tomography (HRCT) in frozen *Betula pendula* branches showed that although the size of gas bubbles increased with conduit size in accordance with the thaw-expansion hypothesis, large bubble-free vessels were often surrounded by small fibers with gas bubbles (Lintunen et al., 2022). Remarkably, only very few gas bubbles were observed in frozen, saturated branches of *Betula pendula*, whereas more bubbles were found in frozen branches with slightly decreased water potential (Lintunen et al., 2022), which is in line with the lower embolism and acoustic emissions recorded in saturated samples (Mayr et al., 2007; Charrier et al., 2014a). Although the mechanism proposed by Charrier *et al.* (2014a) reconciles both freeze-cavitation and thaw-expansion hypothesis, it does not explain how large vessels can be bubble-free while surrounding small fibers contain gas bubbles (Lintunen *et al.*, 2022).

The mechanical constraints, induced by the volume increase due to the water–ice transition (+ 9% compared with liquid water), is strong enough to generate frost cracks (Ishida, 1963; Cinotti, 1991). By generating positive pressure, ice volumetric increase is likely to have a significant impact on the physical state of water during freezing, but this aspect was not explored so far. Furthermore, as bigger bubbles are expected to form in wider vessels at the intra-species or intra-sample scale, wider conduits should therefore embolize first, whereas narrow conduits would embolize only at lower minimum temperature (*i.e.* lower tension; Utsumi *et al.*, 1998; Lemoine *et al.*, 1999; Charrier *et al.*, 2014a). Thus, if wider vessels are preferentially embolized, the mean size of embolized conduits should decrease after each successive freeze-thaw cycle. To explore the role of ice volumetric increase and size of the vessel in embolism formation at a fine spatial and temporal scale, we measured the embolism development by non-destructive visualization using HRCT, which allow direct non-invasive visualization of embolism at the cell level (Cochard *et al.*, 2015; Choat *et al.*, 2016; Nolf *et al.*, 2017). The simultaneous combination of *in vivo* visualization of embolism formation and UE records during freeze-thaw cycles should bring a better understanding of how freeze-thaw induced cavitation, local pressure and embolism occur with respect to xylem anatomy. We therefore designed a dedicated protocol to freeze and thaw samples at a controlled rate while recording tomographic images at high temporal resolution and, in parallel, UEs from samples. This experiment tested the effect of successive freeze-thaw cycles (1 to 3) and minimum temperature (-10 and -20°C) on the formation of embolism in stems of a diffuse-porous angiosperm species (*Betula pendula*). As an accurate measurement of water potential is not feasible in leafless branches, we compared the results at different moment: during the leafy period where stem water potential is measurable, and during leafless period on branches in their natural state. In addition, we visualized gas bubbles formed during ice propagation in *Betula pendula* stems to detect how gas bubbles distribute among vessels of different sizes.

## Material and Methods

### *Plant material*

Branches of *Betula pendula* Roth (approximately 1 m in length, 2 cm in basal diameter corresponding to 2-4 years age) were sampled at different periods on mature trees. For embolism measurement, branches, sampled in the orchard of INRAE Clermont-Ferrand (GPS

45.773 N, 3.144 E) before the first freezing event and leaf fall (November 5$^{th}$), were identified as leafy period samples. Branches, sampled later in the autumn, after leaf fall (November 26$^{th}$) from trees naturally growing in Grenoble (GPS 45.210 N, 5.688 E), were identified as leafless period samples. Additionally, for the distribution of gas bubbles in the frozen sap, 1-m-long branches were collected in early May, soon after budburst, in Southern Finland.

*Embolism measurement*

During the leafy period, the branches were rehydrated over-night the open end immerged in water in a dark room. The branches were moderately dehydrated on the bench until the water potential reached the theoretical value inducing 12% loss of hydraulic conductivity (Charrier *et al.*, 2014). The water potential was measured on end twigs during the dehydration using a Scholander pressure chamber (model 1000 pressure chamber; PMS Instrument). Branches reached a stem water potential of -0.93 ± 0.02 MPa (mean ± SD). Leafless period samples were taken in their natural state as water potential cannot be measured in discontinuous hydraulic systems. Since the beginning of the frost period, the trees were exposed to up to five nights with negative minimal temperature, but higher than -2.2°C.

Branches were cut to approx. 40 cm in length (*ca.* two times the average vessel length) and immediately wrapped in Parafilm (Alcan, Montreal, QC, Canada) to prevent dehydration. The leafy period experiments were carried out at the PIAF laboratory (INRAE, Clermont-Ferrand, France) using a bench X-ray microtomograph (Nanotom 180 XS, GE, Wunstorf, Germany). The other experiments were carried out in Grenoble (France) in the ID19 HRCT beamline of the European Synchrotron Radiation Facility (ESRF).

*Static experiments*

Samples were exposed to freeze-thaw (FT) treatments in a temperature-controlled chamber connected to a circulator bath (Ministat Huber, Offenburg, Germany). Changes in temperature were applied at 5K.h$^{-1}$, and samples were held at target temperature as long as samples were successively scanned (1-3 hours). The temperature in chambers was recorded using thermocouples and a datalogger (CR1000, Campbell, UK). Two different temperature regimes were applied on n = 3 samples (Experiment 1 and 2; Fig. S1):

Experiment 1

Samples were exposed to one FT cycle from +5°C to -20°C and back to +5°C. Samples were scanned at +5°C before FT treatment and after thawing at +5°C. During the leafless period, additional scans were performed during the freezing steps at -10 and -20°C. During the FT treatments, samples were taken out of the temperature-controlled chamber, immediately inserted in a polystyrene insulating cylinder for the X-ray scan (*ca.* 1 minute duration) and put back into the chamber. The total time spent outside the chamber did not exceed 6 minutes and, during the scan, samples did not experience more than 5°C warming (Charra-Vaskou *et al.*, 2016). The leafy samples were only scanned before and after thawing.

Experiment 2

Samples were exposed to three successive FT cycles from +5°C to -10°C and back to +5°C. The leafless samples were scanned five times (frozen and thawed state): at +5°C before FT treatment, after reaching the minimum temperature of -10°C of the first and second cycle (leafless period only), after thawing at +5°C of the first and third cycles. The leafy samples were only scanned three times at +5°C before FT treatment and after thawing at +5°C of the first and third cycles.

*Dynamic experiment*

Two leafless samples were frozen and thawed under the beamline of ESRF. The samples were placed inside a polycarbonate tube on the rotation stage. The temperature was controlled from the control room using a cryostream (Oxford Cryosystems Ltd, UK), which allowed the control of the temperature of the gas flowing out. The temperature change was similar to the previous experiment (5 K.h$^{-1}$) with a minimum temperature set at -15°C. The samples were scanned twenty-three times during the whole FT cycle (*ca.* every 15 minutes). During each scan, X-ray exposure led to local heating in the sample (2-3 cm from the scanned area) of +2 K at a maximum.

During the experiment, ultrasonic acoustic emissions were also monitored on the same samples using a four-channel USB-based system (1283 USB AEnode, 18-bit A/D, 20 MHz) and 150-kHz, 26-dB preamplified resonance sensors (PK15I; all components from Physical Acoustics). Registration and analysis of ultrasonic events were performed with AEwin software version E.4.40 (Mistras Holdings). The sensor was attached using spring-loaded clamp at the base of the sample after removing ca. 1cm² of bark and covering the exposed xylem with silicone grease to ensure acoustic coupling and prevent dehydration. Acoustic coupling was tested with lead breaks (Hsu-Nielsen method; Charrier *et al.*, 2014b) at a distance of 1 cm from the sensor, and sensor was reinstalled when the signal amplitude was below 90 dB. A lead shell was installed around the sensor in order to protect it from the X-ray beam (Fig. S2).

### *Distribution of air bubbles in frozen sap*

Three branches with small leaves were recut (*ca.* 10 cm removed from the base of the 1m-long-branches) under water after sampling, saturated and then bench-dehydrated to a water potential of -0.57 ± 0.05 MPa (mean ± SD). The samples were then frozen in controlled conditions in a test chamber (Weiss Umwelttechnik WK11 2340/40, Vienna, Austria) by first dropping the temperature to +5°C for 15 minutes and then decreasing it to -15°C with constant rate in 1 h 45 minutes. The frozen branches were cut into 80-mm-long samples with a diameter of *ca.* 5 mm, sealed in plastic tubes, stored for a week in a -20°C freezer and then transported frozen in dry ice to the ESRF for HRCT imaging.

In ESRF, the frozen samples were recut to a length of *ca.* 40 mm and placed inside a cryo-durable plastic tube inside a sample holder (see details about the design and experiment in Lintunen et al., 2022) together with dry ice that allowed the sample to stay frozen during the micro-CT imaging. The sample holder was designed with a window in the dry ice in order to allow a direct access to the sample.

**High resolution X-ray computed tomography**

For the leafy period samples, the field of view was set to $5.1\times5.1\times5.1$ mm$^3$ and the X-ray source to 60 kV and 240 µA using a bench X-ray microtomograph (Nanotom 180 XS, GE, Wunstorf, Germany). One thousand images were recorded during the 360° rotation of the sample. The 3D reconstruction was performed using Phoening datosx 2 software (General Electric, Boston, MA, USA) and provided a spatial resolution of 3.8 µm$^3$ per voxel.

For the leafless period samples, we used a 36 keV monochromatic X-ray beam. The projections were recorded with a PCO Dimax, PCO Edge, PCO 4000 camera equipped with a 250 µm thick LuAG scintillator. The complete tomographic scan included 1000 projections, 100 ms seconds each, for a 180° rotation. Tomographic reconstructions were performed using PyHST2 software (Mirone *et al.*, 2014) using the Paganin method (Paganin, 2006), resulting in 2048$^3$ 32-bit volumic images with a final spatial resolution of 6.45 µm$^3$ per voxel.

For the visualization of gas bubbles, a similar set-up as during the leafless period was used with 35 keV X-ray beam. Altogether 15 scans per samples with 0.36 µm pixel size and 0.7 x 0.7 x 0.7 mm$^3$ field of view were recorded from the branches from the center to the surface along a radial gradient.

**Image analysis**

One representative transverse cross section of the 3D tomographic volume was extracted from the middle of the volume for each scan. Air-filled vessels were highly contrasted compared to surrounding tissues. A binary image was thus generated using ImageJ software (http://rsb.info.nih.gov/ij). The diameter and area of each individual air-filled vessel was extracted and measured in the whole cross section, discarding *lumina* with area smaller than 10 µm². The total number and area of air-filled vessels were recorded ($n_{afv}$ and $A_{afv}$, respectively). At the end of the experiments, all samples were cut 2 mm above the previously scanned area, warmed for 5 minutes at 30°C to ensure the thawing and evaporation of the sap in all scanned vessels. A final reference scan, with all vessels air-filled was then recorded. The total number ($n_{tot}$) and area ($A_{tot}$) of all vessels were measured.

The embolized fraction of the vessels ($E_N$) was calculated according to the number of air-filled vessels ($n_{afv}$):

$$E_N = 100 \cdot \frac{n_{afv}}{n_{tot}} \qquad (1)$$

The embolized area fraction of the vessels ($E_A$) was also calculated according to the total area of air-filled vessels ($A_{afv}$):

$$E_A = 100 \cdot \frac{A_{afv}}{A_{tot}} \qquad (2)$$

In addition, the mean hydraulic diameter ($D_h$) was calculated from individual vessels diameter (d) as:

$$D_h = \frac{\sum d^5}{\sum d^4} \qquad (3)$$

The theoretical specific hydraulic conductivity of a cross section ($K_H$) was calculated from the Hagen-Poiseuille equation using the individual diameter of sap- and air-filled vessels as:

$$K_H = \sum \frac{\pi \cdot \emptyset^4}{128 \cdot \eta \cdot A_{Xyl}} \qquad (4)$$

with $K_H$: specific theoretical hydraulic conductivity (kg.m$^{-1}$.MPa$^{-1}$.s$^{-1}$); ⌀: mean feret diameter of vessels (m), η: viscosity of water (1.002 mPa.s at 20°C), and $A_{Xyl}$: xylem area of the cross section (m²).

The theoretical loss of hydraulic conductivity ($E_k$) was calculated as:

$$E_k = 100 \cdot \frac{K_{HA}}{K_{HMax}} \tag{5}$$

with $K_{HA}$ and $K_{HMax}$ representing the theoretical hydraulic conductivities of air-filled vessels, in initial and cut cross sections, respectively.

For the distribution of gas bubbles, the vessels and gas-filled volumes (above 1 µm in diameter) within the vessels were segmented manually from the 3D images. The diameter of each ice-filled vessel was calculated from manually measured cross-sectional area (assuming a spherical cross-section) from one transverse cross section of the 3D tomographic volume from the middle of the volume for each scan (i.e. the middle slice). The total number of vessels examined was 1351.

**Electrolyte leakage**

Cellular damages induced by experimental treatments were measured on the leafy period samples that were not exposed to HRCT to measure the frost hardiness of the samples and on samples that were exposed to HRCT to evaluate the cellular damages induced by FT cycles and HRCT exposure. To measure the frost hardiness, samples were cut into six 5-cm long segments without buds. Samples were exposed to four different freezing temperatures: -10, -20, -30, -40°C. Two control samples were exposed to control (+5°C) and maximal freezing temperature (-75°C). Frozen and thawed rate was set to 5K.h$^{-1}$. Details are provided in Charrier and Améglio (2011). Relative electrolytic leakage (*REL*) was calculated as described in Zhang & Willison (1987). We assumed the following relationship between *REL* and temperature (*θ*) for each sample:

$$REL = \frac{a}{(1+e^{b(c-\theta)})} + d \tag{6}$$

where parameters *a* and *d* define asymptotes of the function, and *b* is the slope at the inflection point *c*.

Frost hardiness and frost sensitivity were estimated as the temperature of the inflection point (c) and the slope at this point (b) and of the adjusted logistic sigmoid function Eq. (6). Parameter estimation was performed by nonlinear regression using nls function in R (R development Core team, 2021).

An index of damage $I_{Dam}$ was calculated after HRCT by measuring REL after the last scan was performed as:

$$I_{Dam} = \frac{REL-d}{a} \qquad (7)$$

where parameters *a* and *d* are the asymptotes of the frost hardiness function.

**Statistical analysis**

Quantitative analysis of data was performed by analysis of variance (ANOVA) after testing the normality of the distribution with Shapiro–Wilk test and homogeneity of variances with Levene test, using R software (R Development Core Team, 2005). For non-normal distribution, a Kruskal–Wallis test was used. The comparisons between vessel diameter distributions was performed using a Chi-square test, corrected for continuity according to Yates. Linear and non-linear regressions (Gaussian function) were performed using the lm and nls functions in R, respectively.

# Results

*Occurrence of embolism after freeze-thaw cycles*

During the leafy period, a moderate increase in the embolism was observed after one freeze-thaw (FT) cycle, either at -10°C ($E_N$ = 23.4 ± 3.8 %; Fig. 1B) or -20°C ($E_N$ = 18.0 ± 4.1 %; Fig. 1F) with no significant effect of the minimum temperature ($F_{1,4}$ = 0.623, $P$ = 0.474; Fig. 1H).

After three FT cycles at -10°C, higher embolism was observed ($E_N$ = 36.8 ± 7.6 %; Fig. 1C), although the difference was not significant ($F_{1,4}$ = 1.686, $P$ = 0.264). The freezing temperature of the xylem sap was recorded at -5.98 ± 0.29°C and no change in the freezing temperature was observed when applying successive FT ($\chi^2$ = 1.308, df = 2, $P$ = 0.520). In parallel, control samples that were maintained at 5°C did not exhibit significant change in the number of embolized vessels for the duration of the experiment (6.8 ± 1.5 and 9.0 ± 1.5 %; before and at the end of experiment, respectively; $F_{1,6}$ = 0.08, $P$ = 0.787; Fig. S3).

During the leafless period, a moderate increase in $E_N$ was observed after one FT cycle to -10°C ($E_N$ = 27.5 ± 3.6 %; $F_{1,4}$ = 19.6, $P$ = 0.011), while higher $E_N$ was observed at lower minimum temperature (*i.e.* -20°C): $E_N$ = 53.5 ± 1.1 % ($F_{1,4}$ = 32.43, $P$ = 0.005). After three successive FT cycles at -10°C, no significant increase in embolism was observed ($E_N$ = 35.4 ± 8.8 %, $F_{1,4}$ = 0.467, $P$ = 0.532; Fig. 1H). Finally, across sampling periods, the initial level of embolism was highly similar: $E_N$ = 7.6 ± 1.3 and 5.8 ± 1.2 % during the leafy and the leafless period, respectively. Only the FT treatment at -20°C was significantly different ($F_{1,4}$ = 47.82, $P$ < 0.001; Fig. 1H).

*Distribution of embolism in vessels*

The mean diameter of the air-filled vessels after FT treatments were not significantly different from the mean diameter of the vessels that were initially embolized ($F_{3,11}$ = 0.279; $P$ = 0.840 and $F_{3,11}$ = 0.291; $P$ = 0.830 in leafy and leafless period, respectively) or from the whole population of vessels ($F_{3,11}$ = 0.291; $P$ = 0.831 and $F_{3,11}$ = 1.439; $P$ = 0.284 in leafy and leafless periods, respectively; Fig. 2A). A similar trend was observed when considering an index that gives higher weight to wide vessels *i.e.* hydraulically-weighed vessel diameters ($D_h$; Fig. S4). Across sampling periods, significant differences were observed for the initial and the whole population of vessels as well as after one FT cycle at -20°C (Fig. 2A). However, significant differences in anatomy between samples analyzed in the two sampling periods were only observed for mean vessel diameter ($F_{1,14}$ = 10.02, $P$ = 0.007) but not for $D_h$ ($F_{1,14}$ = 2.774, $P$ = 0.118), nor theoretical hydraulic conductivity $K_h$ ($F_{1,14}$ = 0.567, $P$ = 0.464).

The correlation between embolism $E_N$ and the mean hydraulic diameter of the whole population of vessels was not significant considering the different treatment individually (1 cycle at -20°C: $F_{1,4}$ = 6.882, $P$ = 0.059; 1 cycle at -10°C: $F_{1,4}$ = 0.034, $P$ = 0.863; 3 cycles at -10°C: $F_{1,5}$ = 0.058, $P$ = 0.819) or together ($F_{3,15}$ = 1.892, $P$ = 0.174). However, the correlation between FT-induced embolism $dE_N$ and the mean hydraulic diameter of embolized vessels was significant considering an effect of the treatment ($F_{3,15}$ = 5.207, $P$ = 0.012; Fig. 2B). Considering the treatments independently, no significant effect of the mean hydraulic diameter of embolized vessels was observed at -10°C ($F_{1,4}$ = 0.042, $P$ = 0.848 and $F_{1,5}$ = 2.836, $P$ = 0.153 for one and three FT cycles, respectively) but at -20°C ($F_{1,4}$ = 16.48, $P$ = 0.015). Between one and three FT cycles at -10°C, no significant effect were observed on the mean diameter of embolized vessels ($F_{1,4}$ = 0.178; $P$ = 0.695 and $F_{1,4}$ = 0.283; P = 0.617 during leafy and leafless period, respectively).

A strong correlation was observed between $E_N$ (air-filled vessels) and $E_A$ (air-filled area; Fig. 2C). Only few points exhibited higher proportion of $E_A$ than of $E_N$ and they were mainly observed in unfrozen samples. In unfrozen samples, the slope of the linear regression between $E_N$ and $E_A$ was higher than 1 (1.177 ± 0.086, mean ± SE) whereas, in the samples that experienced FT cycles, lower than 1 (0.908 ± 0.085, mean ± SE). The slopes were significantly different between unfrozen and frozen-thawed samples (t = 2.214, P = 0.034).

After one or three FT cycles, the distribution of air-filled vessels remained similar to the whole population of the vessels during both the leafy (Fig. 3A) and leafless periods (Fig. 3B). However, it should be noted that although the distribution was significantly different from the whole population of vessels during the leafless period (P < 0.001) and after one FT cycle at -10°C during the leafy period (P = 0.011), the larger vessels did not have higher embolism rate.

The proportion of air-filled vessels per class of diameter was only high in the samples that were exposed to 3 FT cycles during the leafy period. However, in these wider classes, only a few vessels were measured.

*Distribution of embolism and bubbles in frozen stems*

In frozen samples exposed to FT cycles at -10 or -20°C (Fig. 4), $E_N$ remained similar to the initial observations during freezing: $E_N = 5.5 \pm 1.1$ % at -10°C ($\chi^2 = 0.103$, df = 1, $P = 0.749$) and $10.5 \pm 4.6$ % at -20°C ($\chi^2 = 0.429$, df = 1, $P = 0.513$; Fig. 4). After thawing, xylem embolism significantly increased reaching higher values than 50% in the leafless period. Although $E_N$ increased after one FT cycle at -10°C ($E_N = 27.5 \pm 3.6$ %), there was a striking decrease in the next freezing phase, reaching a value of $14.3 \pm 2.2$ %. Finally, after a succession of 3 cycles, a rate of $35.4 \pm 8.8$ % is reached (Fig. 4F).

To better characterize this surprising decrease in embolism rate during freezing, a dynamic visualization of the freezing process was performed during one freeze-thaw cycle with a minimal temperature of -16.5°C. In total, 23 scans were performed while ultrasonic acoustic emissions were continuously recorded (cumulated number of UEs *ca.* 4500; Fig. 5). Although the dynamics were highly similar between growth rings, a twofold higher embolism rate was observed in the external ring compared to the internal ring, at all stages of the freeze-thaw cycles. As long as the temperature was positive (t = 0-2h), $E_N$ remained stable (ca. 15% on average; 20% and 11% in the external and internal rings, respectively) and no UEs were recorded. When the temperature decreased below 0°C, X-ray observation indicated a slight decrease in $E_N$ from 15 to 9% at t = 2-2.5h (11 and 6% in the external and internal rings, respectively). Longitudinal observation of the vessels showed that this decrease in embolism was induced by the entry of ice into air-filled vessels. In particular, some vessels appeared to be filled with a mixture of gas and ice. In parallel, significant acoustic activity was only recorded after $E_N$ stabilized at 9% and before the temperature started to increase (t = 3 - 4.5h).

During thawing, no ultrasonic acoustic emission were recorded, while $E_N$ increased from 9 to 27% (29% and 19% in the external and internal rings respectively) after the temperature became positive (t = 5.5h).

The pattern of gas and ice mixture was explored in frozen samples. The proportion of xylem vessels containing gas bubbles increased with the mean diameter of the vessel, reaching up to 50% in the widest vessels (Fig. 6). The shape of the bubbles was often irregular in the longitudinal direction. In some gas bubbles, visible protrusions of ice towards the center of the lumen of the vessel were observed, while such a pattern was not detectable in fibers (Fig. 7).

*Frost hardiness and vitality of living cells*

In early November, during the leafy period measurements, the frost hardiness of living cells (i.e. temperature inducing 50% relative electrolyte leakage) and the frost sensitivity (the increase in relative electrolyte leakage per degrees close to the frost hardiness temperature) were equal to -29.9 ± 4.2°C and 0.73% °C$^{-1}$, respectively. Based on these two indexes, the FT treatments would have theoretically induced less than 0.1% of cellular damages at -10°C and -20°C in autumn. However, after HRCT exposure, higher levels of damages were measured: 21% ± 21% in control treatment, 72% ± 18% after FT cycles at -10°C and 100 ± 0% after FT cycles at -20°C.

# Discussion

Xylem embolism induced by freeze-thaw cycles has been studied for years (*e.g.* Cochard and Tyree 1990; Sperry and Sullivan, 1992; Sperry *et al.*, 1994; Utsumi *et al.*, 1998), but the underlying mechanisms still remain unclear, mainly because analyses on frozen samples are methodically difficult. New insights into the physiological mechanisms leading to frost-induced embolism have been brought by the use of two non-invasive techniques *i.e.* high resolution computed tomography (HRCT; Charra-Vaskou *et al.*, 2016) and ultrasonic acoustic emission analysis (*e.g.* Charrier *et al.*, 2014a; Kasuga *et al.*, 2015). Our experiments provide the first real-time visualization of frost-induced embolism and a simultaneous use of both methods. The combination of these methods confirmed that two distinct phenomena, namely bubble formation (during freezing) and embolism formation (during thawing) occur within xylem.

In using HRCT to study freeze-thaw process, it should be considered that X-ray exposure can generate deleterious effects on living cells (Petruzelli *et al*., 2018). A high rate of cellular damage was verified in our samples after repeated X-ray exposure, beyond the assumed effects of freezing stress (frost hardiness of -30°C at the time of the measurement). Thus, given the extent of the damage in the area affected by the X-rays, the potential impact of living cells in the freeze-thaw process remains open in our study (Petruzzellis *et al.*, 2018). However, the mechanisms considered in this study are limited to the xylem tissue, whereas the majority of living cells are located in the phloem and cambium. Although living cells can have a role in embolism repair (via stem pressure; Alvès *et al.*, 2004), the embolism formation involves water and gas fluxes between fibers and vessels, non-living elements, on a relatively short time scale. The only process potentially impacted could be the reabsorption of water attracted to ice nucleation points during thawing (Ball *et al.*, 2006). However, this would only affect observations during successive FT cycles. Thus, the potential artifact related to the use of X-rays should not have impacted the main results and conclusions of this study.

The direct observations of vessel content by HRCT suggest that the mechanism of bubble formation in the wood of *Betula pendula* is more complex than previously thought (Pitterman & Sperry, 2003; Charrier et al., 2014; Charra-Vaskou et al., 2016). During the freezing process, the number of air-filled vessels temporarily decreased (Fig. 4). Dynamic observations showed that the ice enters the lumen of the vessels at the moment when acoustic emissions are recorded (Fig. 5). The volumetric increase during the solidification of water is likely to induce water flows such as shown by the ingression of sap in gas filled-vessels (Fig. 7). In vessels, the irregular shapes of the bubbles suggest that liquid fraction of the sap may be submitted to positive pressure from the opposite side (adjacent frozen vessel and/or fiber; Fig. 7). The shape of water ingression into bubbles is similar to the observed shape during drought-induced embolism resorption under positive root pressure (Brodersen *et al.*, 2010; Charrier *et al.*, 2016). Moreover, the shape of the bubbles can also be affected by the temperature gradient, even in the frozen state, since coalescing bubbles were observed after 7 days of storage at -5°C (Carte *et al.*, 1962). In our study, a storage effect may have affected the bubble shape in one of the experiments as samples were stored for one week at -20°C (Fig. 7), but not in the other ones (Fig. 1, 4-5). However, even if the storing conditions in -20C would be considered as part of the freezing treatment, it would not change the main conclusions made from the data.

Would the ingression of water into the vessel lumen be enough to induce mechanical constraint and possibly micro-fractures in the cell walls and thus generate acoustic emission? This hypothesis would only explain the first stage of the acoustic emissions, when the ice nucleates and propagates in the wood (Charrier *et al.*, 2015) but not the emission dynamics at much lower temperatures or the differences between species in emission patterns (Charrier *et al.*, 2014a). Another source of acoustic emissions could be the frost-induced cell dehydration (Charrier *et al.*, 2014a; Kasuga *et al.*, 2016) as shown by Lamacque *et al.* (2022) during drought stress. Cell dehydration, induced by the very low chemical potential of the ice (Mazur, 1969), may have provided an additional emission source increasing the migration of extracellular water to the non-living parts of the xylem. To determine the possible causes of UEs, a dynamic observation at finer scale would be needed to observe pit structure, cell turgor and eventually intracellular cavitation during freezing stress.

Bubbles are only observed in samples under moderate tension (Lintunen *et al.*, 2022) as saturated branches would not allow pit aspiration and bubbles to expand on freezing. In our study, the samples were under moderate tension, either in their natural state (leafless period) or at the water potential inducing 12% loss of hydraulic conductivity (leafy period). Although the exact water potential during the leafless period cannot be measured, the initial level of embolism was similar and consistent with the measured water potential during the leafy period. Samples were covered with parafilm to avoid dehydration during the experimental treatments. This also prevented gas efflux from the sample during freezing and thus potentially increased the concentration of gas in the sap and disrupted axial gas flows (Lintunen *et al.* 2014; 2020). In the dynamic experiment, the outer growth ring had a higher embolism rate than the inner growth ring, but the difference was stable throughout the freeze-thaw cycle, suggesting that there was no consistent radial gas flow.

The freeze-thaw treatments induced substantial and various amount of embolism in the samples, from 10 to 50% allowing to study the vulnerability of individual vessels to FT treatments (Fig. 4). Our results evidenced that, within a sample, wider vessels were not more vulnerable to freeze-thaw induced embolism, although more gas was dissolved in their sap volume and most of these vessels contained gas bubbles. A significant increase in embolism in the largest vessels was only observed during the leafy period, after 3 successive FT cycles. According to the thaw-expansion hypothesis, bigger bubbles in wide vessels would expand at lower tension on thawing (Sperry & Sullivan, 1992). The correlations between the hydraulic diameter of the vessels and the vulnerability to frost-induced embolism have been performed in previous literature typically by interspecific comparisons (Davis *et al.*, 1999; Pitterman and Sperry, 2003; 2006; Charrier *et al.*, 2013, 2014a). At an intraspecific or intra-stem scale, higher embolism in larger conduits has also been observed when comparing protoxylem *versus* metaxylem (Cobb *et al.* 2007), or early *versus* late wood (Ellmore and Ewers 1986; Cochard & Tyree, 1990; Utsumi *et al.*, 1996). In our study, the distribution of diameters of air-filled vessels were similar after successive freeze-thaw cycles, and not significantly different from that of the total population of vessels (Fig. 2, 3). Furthermore, the slope of the regression between $E_N$ and

$E_A$, being very close to 1, indicates that the number of air-filled vessels increased in parallel with the air-filled lumen area (Fig. 2C), suggesting that the previous observations were only correlative and not causal. The large presence of gas bubbles in the fibers that are small in size also does not support any clear relation between the volume of the xylem element and the presence of gas (Lintunen *et al.*, 2022). During freezing, the complex pressure / tension dynamics inside a large vessel in interaction with its surrounding tissues could compensate for the effect of large sap volume on the diameter of the bubble and thus the tension required for it to expand during thawing. However, exploration across a wider range of species would be needed to verify that this is not a mechanism specific to *Betula pendula* although similar observations were made in a vesselless species (*Picea abies*; Mayr *et al.*, 2007).

Differences across sampling periods were only observed after one FT at -20°C (Fig. 1). During the leafless period, the hydraulic connection is interrupted, isolating parts of the wood and making water distribution more heterogeneous, with local xylem area showing gas-filled fibers and more prone to embolism formation. These gas-filled fibers can be a source for embolism propagation but only at relatively low temperature (*i.e.* -20°C).

After each freeze-thaw cycles, water is redistributed, concentrating in the cambium and phloem where the living cells reabsorb water (Fig. 4; Charra-Vaskou *et al.*, 2016). Furthermore, air seeding through pits would allow air bubbles to invade connected vessels (Mayr *et al.*, 2007). These two factors explain why embolism is increasing with the number of freeze-thaw cycles. Although most of embolism was generated during the first cycle, it slightly increased after successive cycles as it has been reported in previous studies (Sperry *et al.*, 1994; Sparks *et al.*, 2001; Mayr *et al.*, 2003; 2007). As previously indicated, the damages caused to living cells by HRCT may have affected the dynamics of water reabsorption in the cambium zone and the accumulation of embolism after several FT cycles.

This work presents the first dynamic observation of freeze-thaw induced embolism using synchrotron-based high resolution X-ray computed tomography, combined with ultrasonic emission analysis. It clearly proves that bubble formation and embolism are two distinct phenomena, occurring at different stages during freeze-thaw cycles. Furthermore, the thaw-expansion is consistent with the observations although it has been challenged in this study by the fact that wider xylem elements do not embolize first within tissue. The size of the individual conduit does not determine the risk for it to embolize within the tissue, although the relation exists at the scale of a branch. This observation questions the underlying mechanistic assumptions that have led to the use of vessel diameter to explain plant species distribution. The physical complexity of the wood during the freezing process (*i.e.* involving water under three physical states, with local pressure and tension) opens many research questions. To better understand the physical mechanism leading to embolism formation, an analysis of cavitation events at a fine spatial scale (*e.g.* within pits, in the lumen of the xylem elements and in living cells) and knowledge on local sap pressure and temperature would be required. Direct *in situ* observations, using high temporal and spatial resolutions of X-ray tomography measurements, as well as an accurate control of the temperature, would thus help to visualize the bubble formation, migration and expansion. A better understanding of gas-water-ice dynamics during freeze-thaw cycles would thus allow to unravel how the hydraulic constraints drive plant distribution in high elevation and latitude.

## Author Contribution

K.C.-V., G.C., A.L. and E.B conceived the original screening and research plans (High Resolution Computed Tomography) with inputs from T.A., S.M. and H.C.. G.C., E.B., S.M., T.A., Y.S., A.L. and H.S. performed the HRCT scans, K.C.-V., G.C. and H.S. analyzed the data K.C.-V. and G.C. wrote the article with contributions of all the authors.

## Conflict of Interest statement

The authors have no conflicts to declare.

## Funding statement

This work was supported by the Agence Nationale de la Recherche (ANR-20-CE91-0008) and the Austrian Science Fund (I 4918-B) through the Acoufollow project, by the Auvergne-Rhone-Alpes council (Pack Ambition Recherche Doux- Glace to G.C. and M.V.R.) and the European Radiation Synchrotron Facility (proposal LS-2243).

## Acknowledgements

We thank Pierre Conchon, Christophe Serre and Romain Souchal for their technical support with the sampling and temperature treatments. We also acknowledge the European Synchrotron Radiation Facilities and the ID19 beamline staff Alexander Rack, Elodie Boller and Lukas Helfen for provision of synchrotron radiation facilities and their technical support for anatomy measurements.

## Data availability statement

Data supporting this study will be available on demand to the corresponding author upon reasonable request.

# Supporting information

**Figure S1.** Temperature courses during the different freeze-thaw experiments. This experimental set-up allowed to test different conditions i.e. the effect of (i) freezing temperature (1 vs 2), (ii) minimum temperature of a freeze-thaw cycle (3 vs 4), (iii) successive freeze-thaw cycles during freezing (1 vs 5), or after thawing (3 vs 6). All stages were performed during leafless period, whereas, during leafy periods, only stage 0, 3, 4 and 6 were performed.

**Figure S2.** Experimental setup used to freeze and thaw samples in the experimental ID19 beamline for combined HRCT imaging and ultrasonic emission analysis. The cryostream device bring cold from the top of the sample holder. The acoustic sensor is plugged at the bottom part of the sample and is protected from the X-ray beam by a lead sheet.

**Figure S3.** Transverse HRCT images and insets at higher magnification of the same stems stored at +5°C in *Betula pendula* during the leafy period before (A) and after being stored between 72 and 96 hours at 5°C (B). At the end of the experiment, the sample was cut above the scan area to visualize all the vessels (C). Bars = 1 mm.

**Figure S4.** Mean hydraulically weighed diameter of air-filled vessels in *Betula pendula* (before and after freeze-thaw cycles: 1 and 3 cycles to -10°C (1FT -10, 3FT -10, respectively) and 1 cycle to -20°C (1FT -20). Symbols represent significant statistical differences between the leafy and the leafless periods (black and white bars, respectively): ns >0.05; * >0.01; ** >0.001; *** <0.001. Letters refer to statistically significant differences across treatments. Lower and major cases for leafy and leafless periods, respectively.

# Figure captions

**Figure 1.** Transverse HRCT images and insets at higher magnification of the same stems acquired during freeze-thaw cycles in *Betula pendula* (A-D) during the leafy period and (E-G) during leafless period, before (A, E), after freezing at -10°C and thawing at 5°C once (B) and after 3 three FT cycles at -10°C (C) and after freezing at -20°C and thawing at 5°C once (F). After the experiment, the sample was cut above the scan area to visualize all the vessels (D, G). Bars = 1 mm. **H.** Percent of air-filled vessels in stems of *Betula pendula* before and after freeze-thaw cycles: 1 and 3 cycles to -10°C (1FT -10, 1FT -10, respectively) and 1 cycle to -20°C (1FT -20). Symbols represent significant statistical differences between the leafy and the leafless periods (black and white bars, respectively): ns >0.05; * >0.01; ** >0.001; *** <0.001. Letters refer to statistically significant differences across treatments. Lower and major cases for leafy and leafless periods, respectively.

**Figure 2. A.** Mean diameter of air-filled vessels in *Betula pendula* before and after freeze-thaw cycles: 1 and 3 cycles to -10°C (1FT -10, 3FT -10, respectively) and 1 cycle to -20°C (1FT -20). Symbols represent significant statistical differences between the leafy and the leafless periods (black and white bars, respectively): ns >0.05; * >0.01; ** >0.001; *** <0.001. Letters refer to statistically significant differences across treatments. Lower and major cases for leafy and leafless periods, respectively. **B.** Increase in embolism $E_N$ depending on the mean hydraulic diameter of the stem of *Betula pendula*. The correlation was only significant for one FT cycle to - -20°C (closed symbols) but not for one or three FT cycles to -10°C (open and gray symbols, respectively). **C.** Percent lumen area depending on number of air filled vessels in unfrozen samples (open symbols) and samples that experience one or several FT cycles (gray symbols) in *Betula pendula*. The points above the y = x line indicates that the proportion of larger vessels in the total number of air-filled vessels is higher.

**Figure 3.** Distribution of air-filled vessels after different freeze-thaw treatments (solid bars) and in the whole section (empty bars) during the leafy (A) and the leafless period (B) in one year stems of *Betula pendula*. The insets represent the ratio of air-filled vessels compared to the total number of vessels in each diameter class for the different freeze-thaw treatments. Note that the primary and secondary y-scales differ.

**Figure 4.** Transverse HRCT images and insets at higher magnification of the same stems acquired during freeze-thaw cycles in *Betula pendula* (A-D) during the leafless period before (A), during freezing at -10°C (B) and -20°C (C) and after thawing at 5°C (D). After the experiment, the sample was cut above the scan area to visualize all the vessels. Bars = 1 mm. **E-F** Percent of air-filled vessels during freeze-thaw cycles in *Betula pendula* during the leafless period. **E** Before freezing, during freezing at -10°C (F-10) and -20°C (F-20) and after thawing **F** Before freezing, during freezing at -10°C (F-10), after 1 FT cycle at thawing (1FT-10) and frozen again (1FT-10F) and completely thawed after 3 FT cycles (3FT -10). Letters refer to statistically significant differences across treatments. The values from thawed samples (as in figure 1) are represented by open bars, whereas frozen samples by gray bars.

**Figure 5.** Transverse HRCT images during freezing and thawing a stem of *Betula pendula* during the leafless period: before freezing (A), during freezing (B) and after thawing (C). Insets represent a zoom on radial and longitudinal views. Bars = 1 mm. Lower panel (D) represents sample temperature, percentage of air-filled vessels and cumulated ultrasonic acoustic emissions during the freeze-thaw cycle.

**Figure 6.** Distribution of vessels in classes of different conduit radius in *Betula pendula*. Insets represent the ratio of vessels with gas bubble(s) to the total number of vessels in each diameter class (note that fully embolized fibers/vessels are not included in the analysis).

**Figure 7.** Longitudinal-radial images of gas voids in frozen *Betula pendula* vessels. Ice is represented in light gray (white arrows) and gas bubbles in dark gray (black arrows).

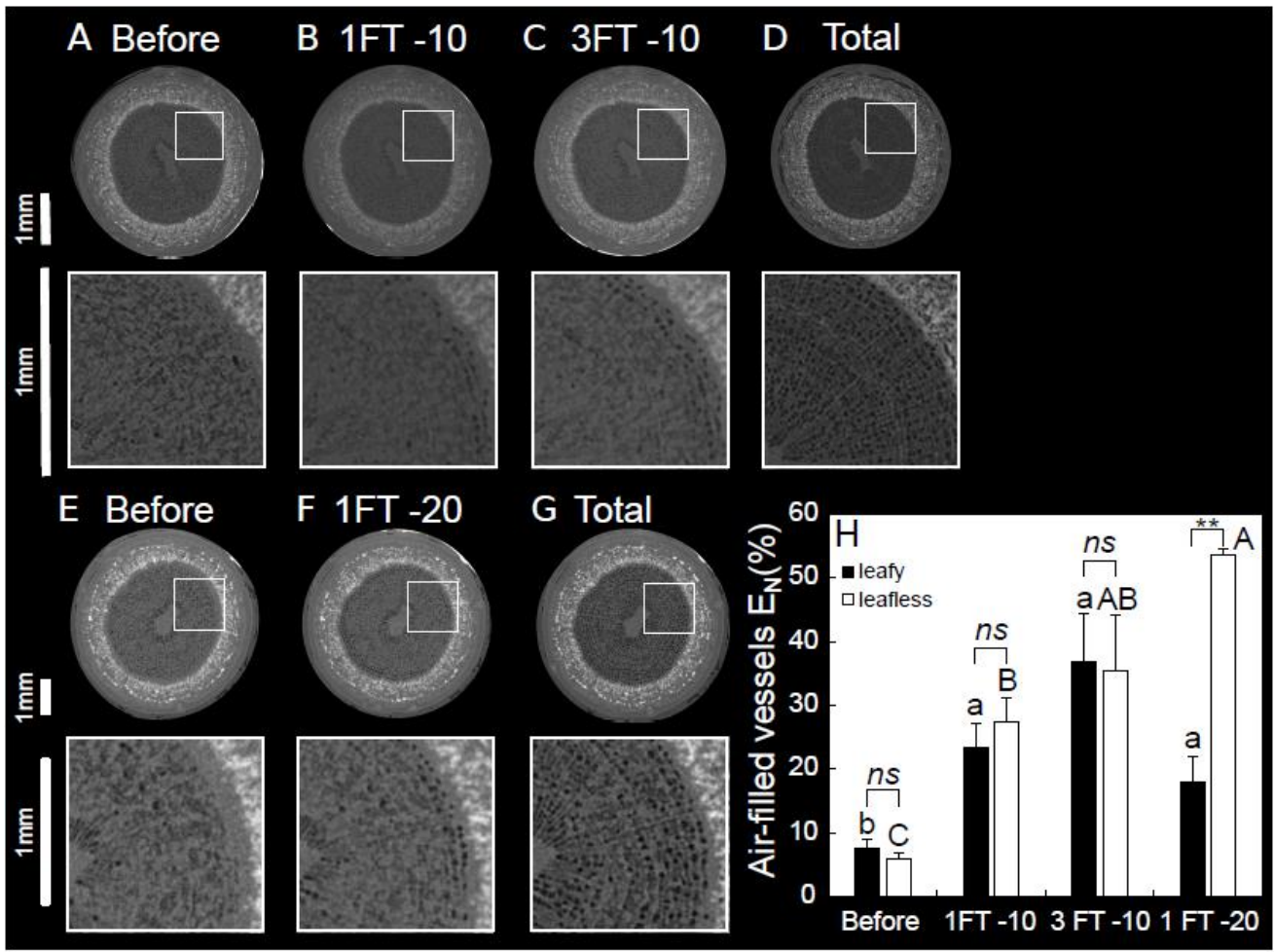

**Figure 1.** Transverse HRCT images and insets at higher magnification of the same stems acquired during freeze-thaw cycles in *Betula pendula* (A-D) during the leafy period and (E-G) during leafless period, before (A, E), after freezing at -10°C and thawing at 5°C once (B) and after 3 three FT cycles at -10°C (C) and after freezing at -20°C and thawing at 5°C once (F). After the experiment, the sample was cut above the scan area to visualize all the vessels (D, G). Bars = 1 mm. **H.** Percent of air-filled vessels in stems of *Betula pendula* before and after freeze-thaw cycles: 1 and 3 cycles to -10°C (1FT -10, 1FT -10, respectively) and 1 cycle to -20°C (1FT -20). Symbols represent significant statistical differences between the leafy and the leafless periods (black and white bars, respectively): ns >0.05; * >0.01; ** >0.001; *** <0.001. Letters refer to statistically significant differences across treatments. Lower and major cases for leafy and leafless periods, respectively.

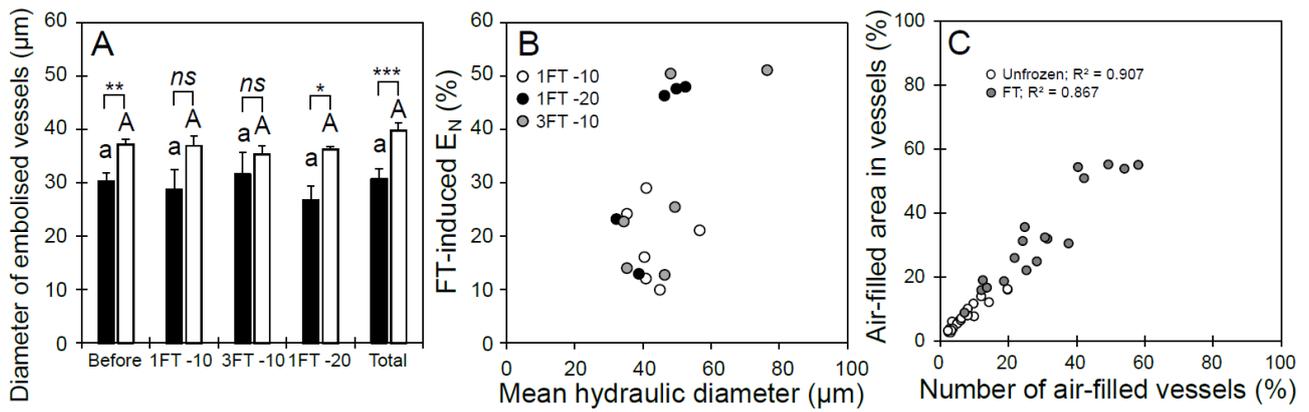

**Figure 2. A.** Mean diameter of air-filled vessels in *Betula pendula* before and after freeze-thaw cycles: 1 and 3 cycles to -10°C (1FT -10, 3FT -10, respectively) and 1 cycle to -20°C (1FT -20). Symbols represent significant statistical differences between the leafy and the leafless periods (black and white bars, respectively): ns >0.05; * >0.01; ** >0.001; *** <0.001. Letters refer to statistically significant differences across treatments. Lower and major cases for leafy and leafless periods, respectively. **B.** Increase in embolism $E_N$ depending on the mean hydraulic diameter of the stem of *Betula pendula*. The correlation was only significant for one FT cycle to - -20°C (closed symbols) but not for one or three FT cycles to -10°C (open and gray symbols, respectively). **C.** Percent lumen area depending on number of air filled vessels in unfrozen samples (open symbols) and samples that experience one or several FT cycles (gray symbols) in *Betula pendula*. The points above the y = x line indicates that the proportion of larger vessels in the total number of air-filled vessels is higher.

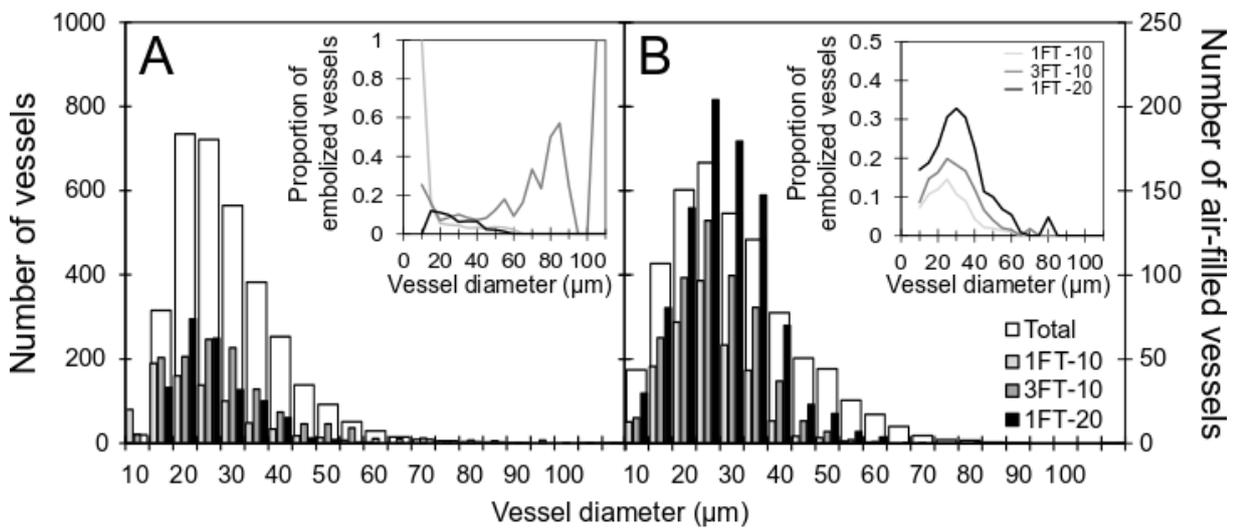

**Figure 3.** Distribution of air-filled vessels after different freeze-thaw treatments (solid bars) and in the whole section (empty bars) during the leafy (A) and the leafless period (B) in one year stems of *Betula pendula*. The insets represent the ratio of air-filled vessels compared to the total number of vessels in each diameter class for the different freeze-thaw treatments. Note that the primary and secondary y-scales differ.

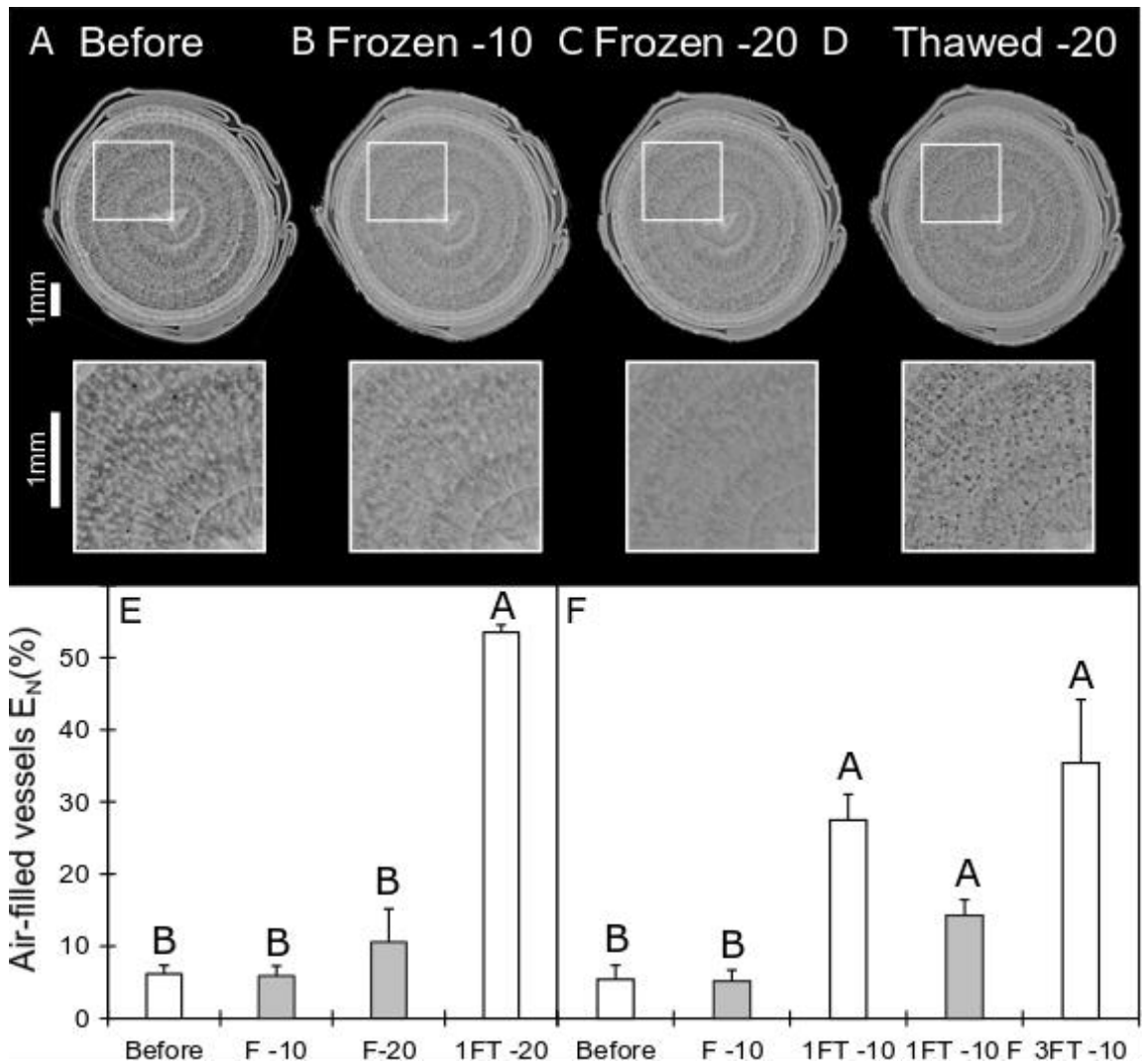

**Figure 4.** Transverse HRCT images and insets at higher magnification of the same stems acquired during freeze-thaw cycles in *Betula pendula* (A-D) during the leafless period before (A), during freezing at -10°C (B) and -20°C (C) and after thawing at 5°C (D). After the experiment, the sample was cut above the scan area to visualize all the vessels. Bars = 1 mm. **E-F** Percent of air-filled vessels during freeze-thaw cycles in *Betula pendula* during the leafless period. **E** Before freezing, during freezing at -10°C (F-10) and -20°C (F-20) and after thawing **F** Before freezing, during freezing at -10°C (F-10), after 1 FT cycle at thawing (1FT-10) and frozen again (1FT-10F) and completely thawed after 3 FT cycles (3FT -10). Letters refer to statistically significant differences across treatments. The values from thawed samples (as in figure 1) are represented by open bars, whereas frozen samples by gray bars.

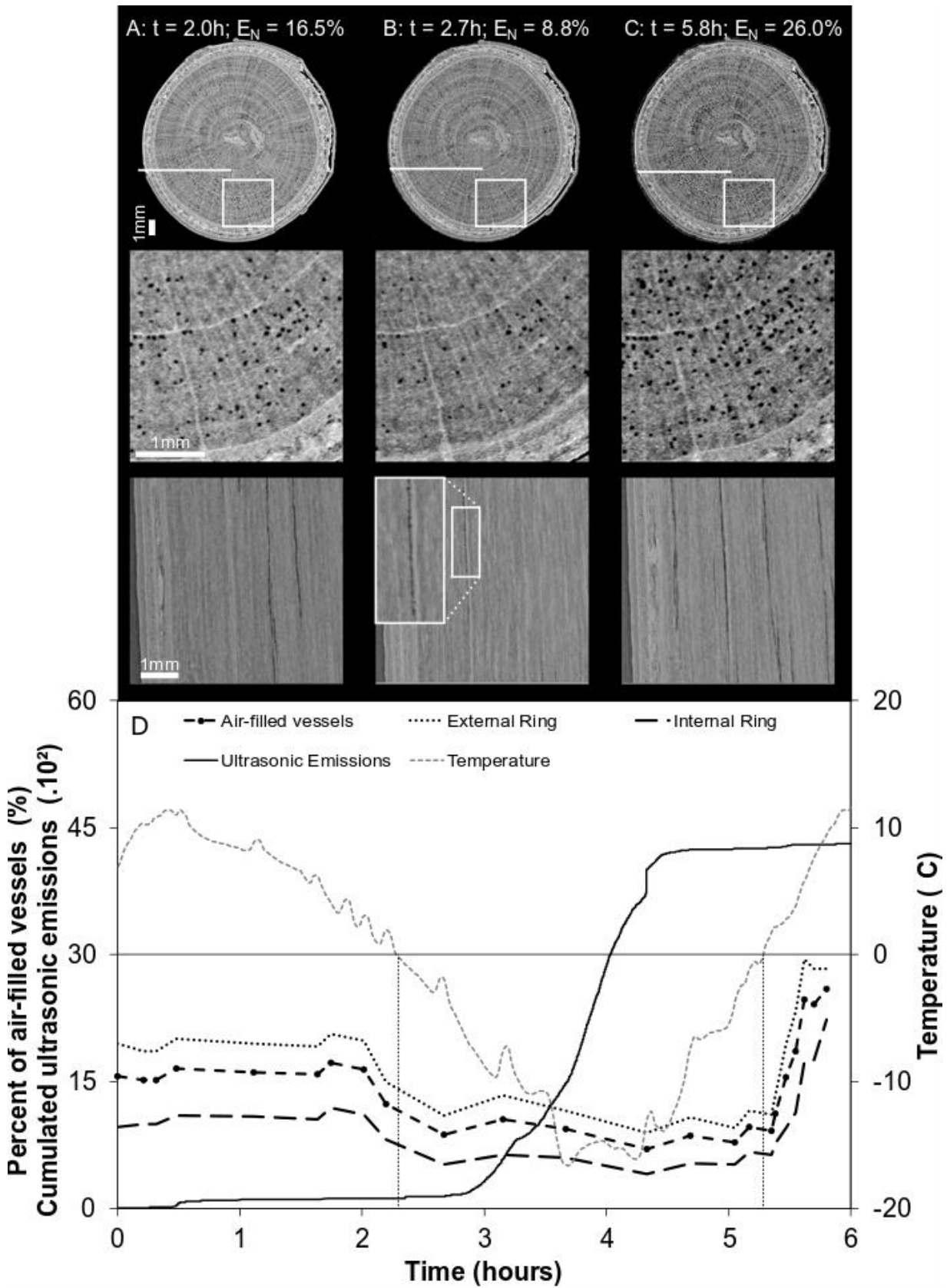

**Figure 5.** Transverse HRCT images during freezing and thawing a stem of *Betula pendula* during the leafless period: before freezing (A), during freezing (B) and after thawing (C). Insets represent a zoom on radial and longitudinal views. Bars = 1 mm. Lower panel (D) represents sample temperature, percentage of air-filled vessels and cumulated ultrasonic acoustic emissions during the freeze-thaw cycle.

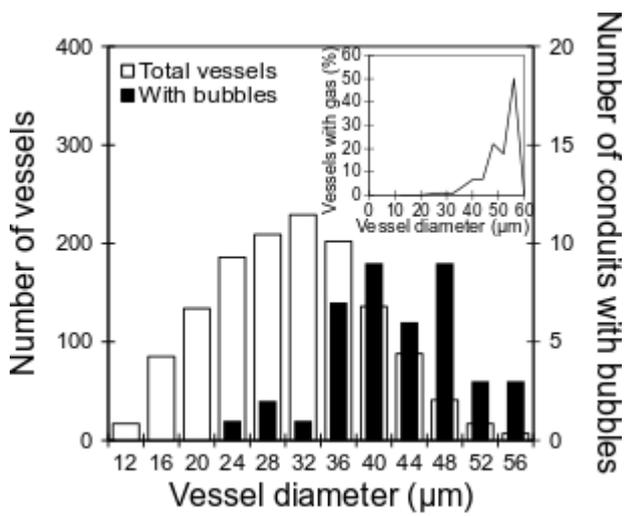

**Figure 6.** Distribution of vessels in classes of different conduit radius in *Betula pendula*. Insets represent the ratio of vessels with gas bubble(s) to the total number of vessels in each diameter class (note that fully embolized fibers/vessels are not included in the analysis).

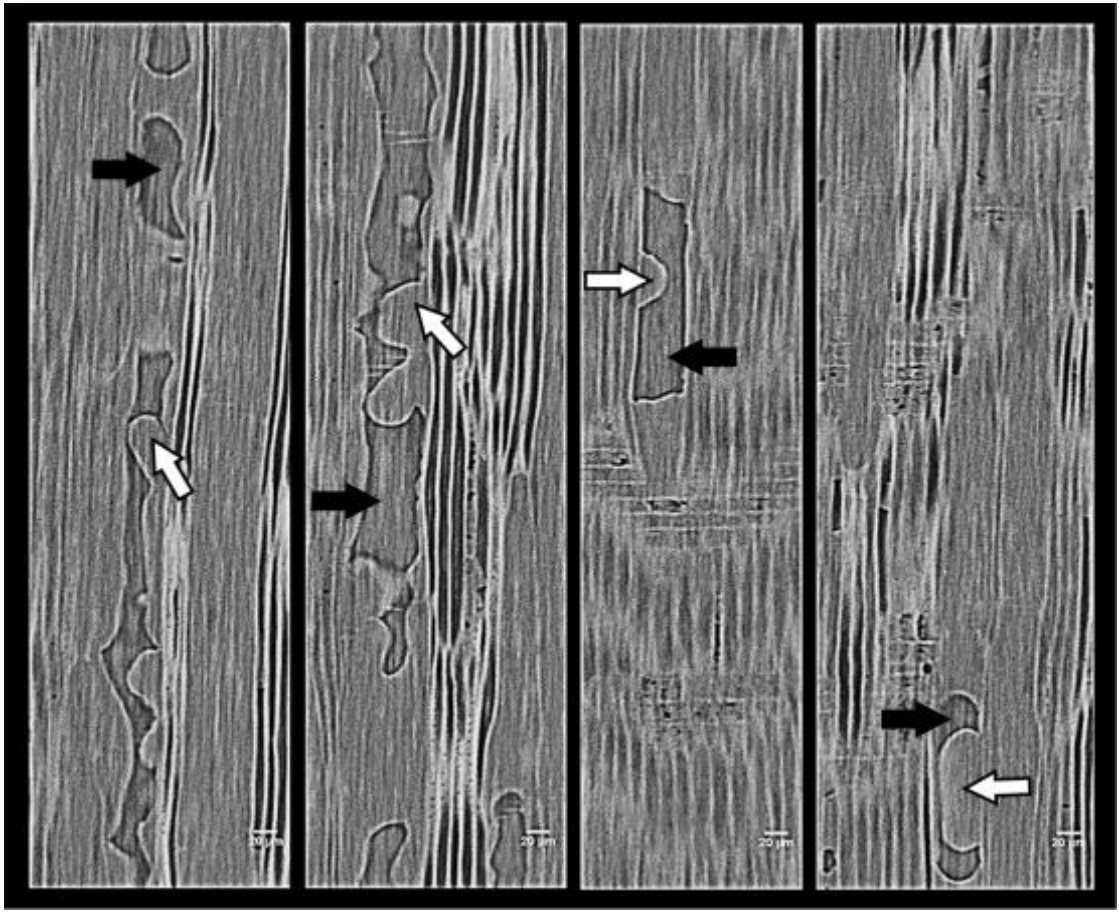

**Figure 7.** Longitudinal-radial images of gas voids in frozen *Betula pendula* vessels. Ice is represented in light gray (white arrows) and gas bubbles in dark gray (black arrows).

**Supplementary material**

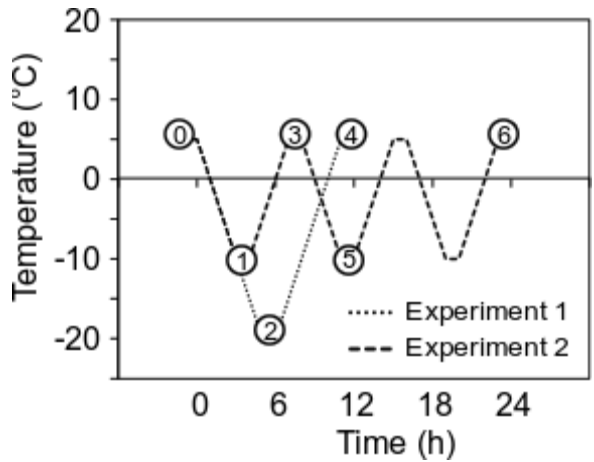

**Figure S1.** Temperature courses during the different freeze-thaw experiments. This experimental set-up allowed to test different conditions i.e. the effect of (i) freezing temperature (1 vs 2), (ii) minimum temperature of a freeze-thaw cycle (3 vs 4), (iii) successive freeze-thaw cycles during freezing (1 vs 5), or after thawing (3 vs 6). All stages were performed during leafless period, whereas, during leafy periods, only stage 0, 3, 4 and 6 were performed.

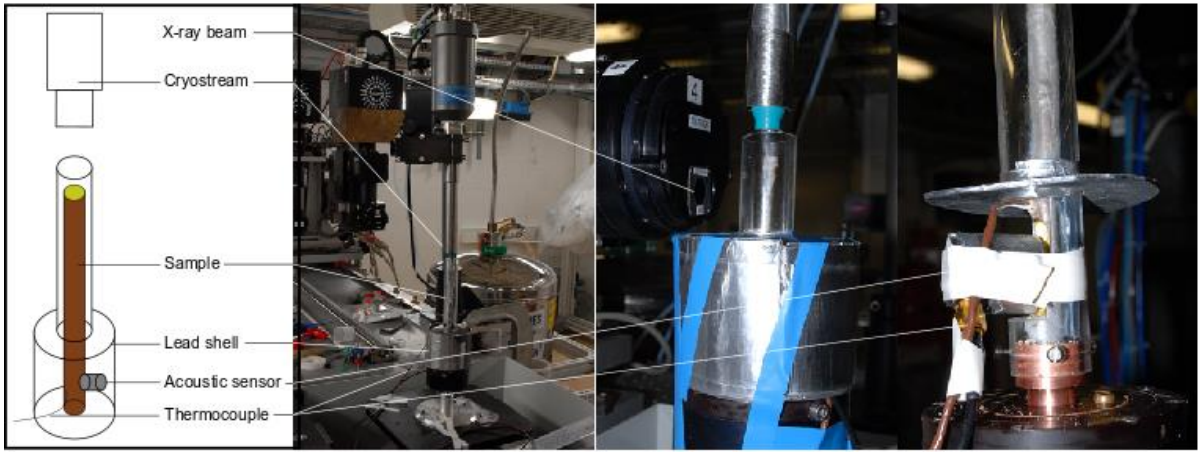

**Figure S2.** Experimental setup used to freeze and thaw samples in the experimental ID19 beamline for combined HRCT imaging and ultrasonic emission analysis. The cryostream device bring cold from the top of the sample holder. The acoustic sensor is plugged at the bottom part of the sample and is protected from the X-ray beam by a lead sheet.

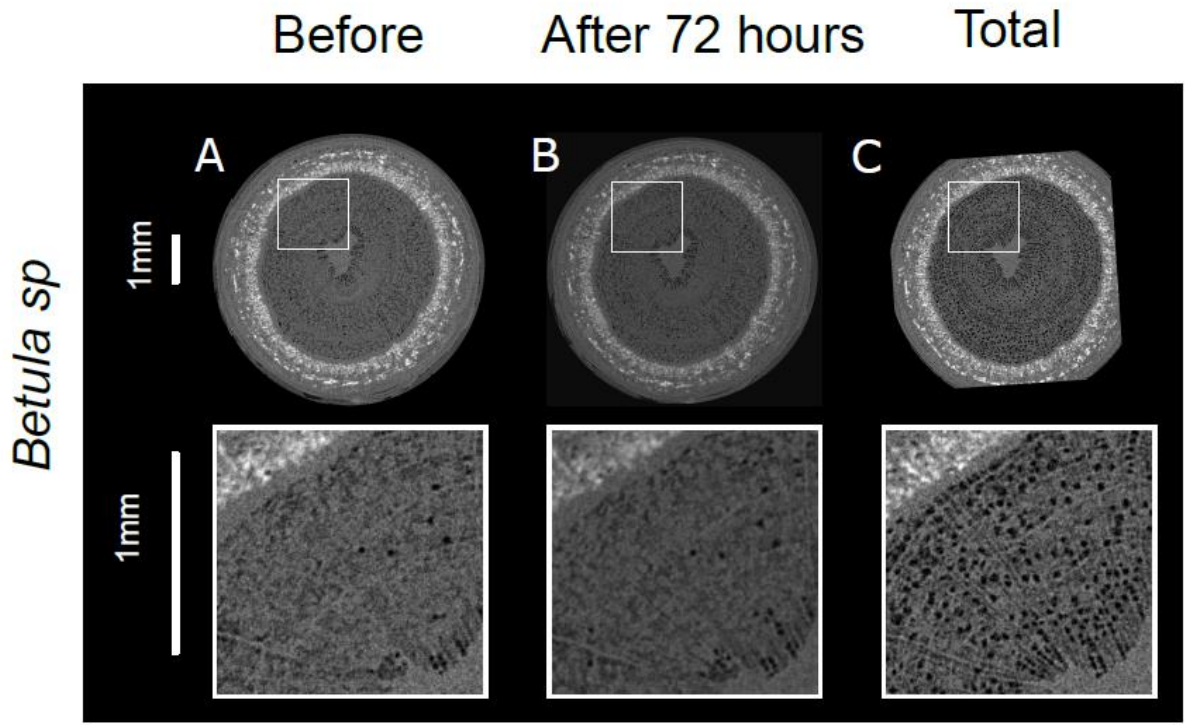

**Figure S3.** Transverse HRCT images and insets at higher magnification of the same stems stored at +5°C in *Betula pendula* during the leafy period before (A) and after being stored between 72 and 96 hours at 5°C (B). At the end of the experiment, the sample was cut above the scan area to visualize all the vessels (C). Bars = 1 mm.

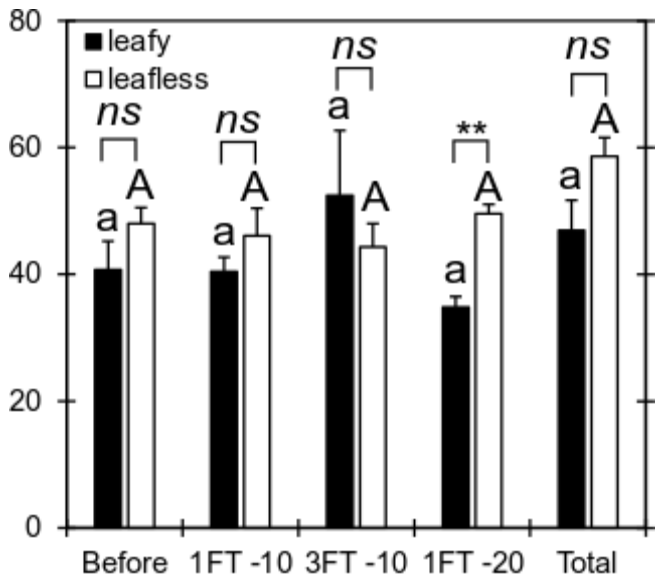

**Figure S4.** Mean hydraulically weighed diameter of air-filled vessels in *Betula sp* (before and after freeze-thaw cycles: 1 and 3 cycles to -10°C (1FT -10, 3FT -10, respectively) and 1 cycle to -20°C (1FT -20). Symbols represent significant statistical differences between the leafy and the leafless periods (black and white bars, respectively): ns >0.05; * >0.01; ** >0.001; *** <0.001. Letters refer to statistically significant differences across treatments. Lower and major cases for leafy and leafless periods, respectively.